\documentclass[floatfix,aip,reprint]{revtex4-1}

\usepackage[utf8]{inputenc}
\usepackage[T1]{fontenc}
\usepackage{textcomp}
\usepackage[euler]{textgreek}
\usepackage{gensymb}
\usepackage[caption=false]{subfig}
\usepackage[section]{placeins}
\usepackage{graphicx}
\usepackage{siunitx}
\usepackage{paralist}
\usepackage{booktabs}
\usepackage{miller}
\usepackage[version=3]{mhchem}
\usepackage{color}
\usepackage{pst-all}
\graphicspath{{../figs/}}

\begin{document}

\hyphenation{pseu-do-mor-phic InGaN GaN ca-tho-do-lu-mi-nes-cen-ce se-mi-bulk tri-me-thyl-in-di-um tri-me-thyl-gal-li-um}

\title{Scalable control of graphene growth on 4H-SiC C-face using decomposing silicon nitride masks.}

\author{Renaud Puybaret}\email{renaudpuybaret@gmail.com}
\affiliation{School of Electrical and Computer Engineering, Georgia Institute of Technology, 30332, Atlanta, Georgia, USA}
\affiliation{Georgia Tech - CNRS UMI 2958, 2-3 Rue Marconi, 57070, Metz, France}

\author{John Hankinson}
\affiliation{School of Physics, Georgia Institute of Technology, 30332, Atlanta, Georgia, USA}

\author{James Palmer}
\affiliation{School of Physics, Georgia Institute of Technology, 30332, Atlanta, Georgia, USA}

\author{Cl\'ement Bouvier}
\affiliation{School of Physics, Georgia Institute of Technology, 30332, Atlanta, Georgia, USA}

\author{Abdallah Ougazzaden}
\affiliation{School of Electrical and Computer Engineering, Georgia Institute of Technology, 30332, Atlanta, Georgia, USA}
\affiliation{Georgia Tech - CNRS UMI 2958, 2-3 Rue Marconi, 57070, Metz, France}

\author{Paul L. Voss}
\affiliation{School of Electrical and Computer Engineering, Georgia Institute of Technology, 30332, Atlanta, Georgia, USA}
\affiliation{Georgia Tech - CNRS UMI 2958, 2-3 Rue Marconi, 57070, Metz, France}

\author{Claire Berger}
\affiliation{School of Physics, Georgia Institute of Technology, 30332, Atlanta, Georgia, USA}
\affiliation{CNRS-Institute Néel, BP 166, 38042 Grenoble, Cedex 9, France}

\author{Walt A. de Heer}
\affiliation{School of Physics, Georgia Institute of Technology, 30332, Atlanta, Georgia, USA}  

\date{November 10, 2014}

\begin{abstract}

{Patterning of graphene is key for device fabrication. We report a way to increase or reduce the number of layers in epitaxial graphene grown on the C-face (000\=1) of silicon carbide by the deposition of a 120 nm to 150nm-thick silicon nitride (SiN) mask prior to graphitization. In this process we find that areas covered by a Si-rich SiN mask have one to four more layers than non-masked areas. Conversely N-rich SiN decreases the thickness by three layers. In both cases the mask decomposes before graphitization is completed. Graphene grown in masked areas show good quality as observed by Raman spectroscopy, atomic force microscopy (AFM) and transport data. By tailoring the growth parameters selective graphene growth and sub-micron patterns have been obtained.}

\end{abstract}

\maketitle


Epitaxial graphene (EG) on SiC has great potential for electronics, with compelling physical characteristics such as ballistic transport in nanoribbons \cite{deheer2014}, half-eV band-gap structures \cite{hicks2012}, highly-efficient spin transport \cite{martin2012}, and ultra-high frequency transistors \cite{avouris2010,guo2013}. Because SiC is a monocrystalline semiconducting industrial substrate, epitaxial graphene on SiC  is directly compatible with established scalable device fabrication techniques, making it attractive for  advanced electronic devices\cite{berger2004,lin2011}. Patterning of graphene devices is a key step in the fabrication process. In most cases, 2D graphene is first grown then patterned by oxygen plasma. Selective area growth is a more straightforward approach, as it could in principle provide shaped structures in a single-step process,  without relying on the usual post-growth lithography-and-etching steps. Several techniques have been reported for local control of graphene growth. These include in particular AlN capping\cite{zaman2010}, ion implantation of Au or Si\cite{tongay2012}, the use of amorphous carbon corrals\cite{palmer2014}, and side-wall nanoribbons\cite{sprinkle2010}.

In this paper, we report a method for controlling graphene growth selectivity down to the sub-micron level in a one-step process with a vanishing mask.
We find that deposition of a 120 nm- to 150 nm-thick silicon nitride mask on C-face (000\=1) 4H-SiC prior to graphitization modifies the relative number of multi-layer epitaxial graphene (MEG) sheets. The silicon nitride mask decomposes and vanishes before graphitization is complete. Importantly, the stoichiometry of the silicon nitride layers controls whether the silicon nitride layer enhances or suppresses graphene growth relative to uncovered areas. We find that N-rich silicon nitride masks decrease the average number of layers by three compared to uncovered regions while Si-rich silicon nitride masks increase thickness by two to four layers. The graphene layers of samples prepared with nearly stoichiometric silicon nitride show good mobilities up to 7100 cm$^2$.V$^{-1}$.s$^{-1}$, with electron concentrations in the 10$^{12}$ cm$^{-2}$ range. Raman spectroscopy and AFM measurements confirm that the graphene grown in areas initially covered by the mask has good structural quality.

\begin{figure}[h]
\centering
\vspace{-0pt}
\begin{center}$
\begin{array}{cc}
\includegraphics[width=.8\linewidth]{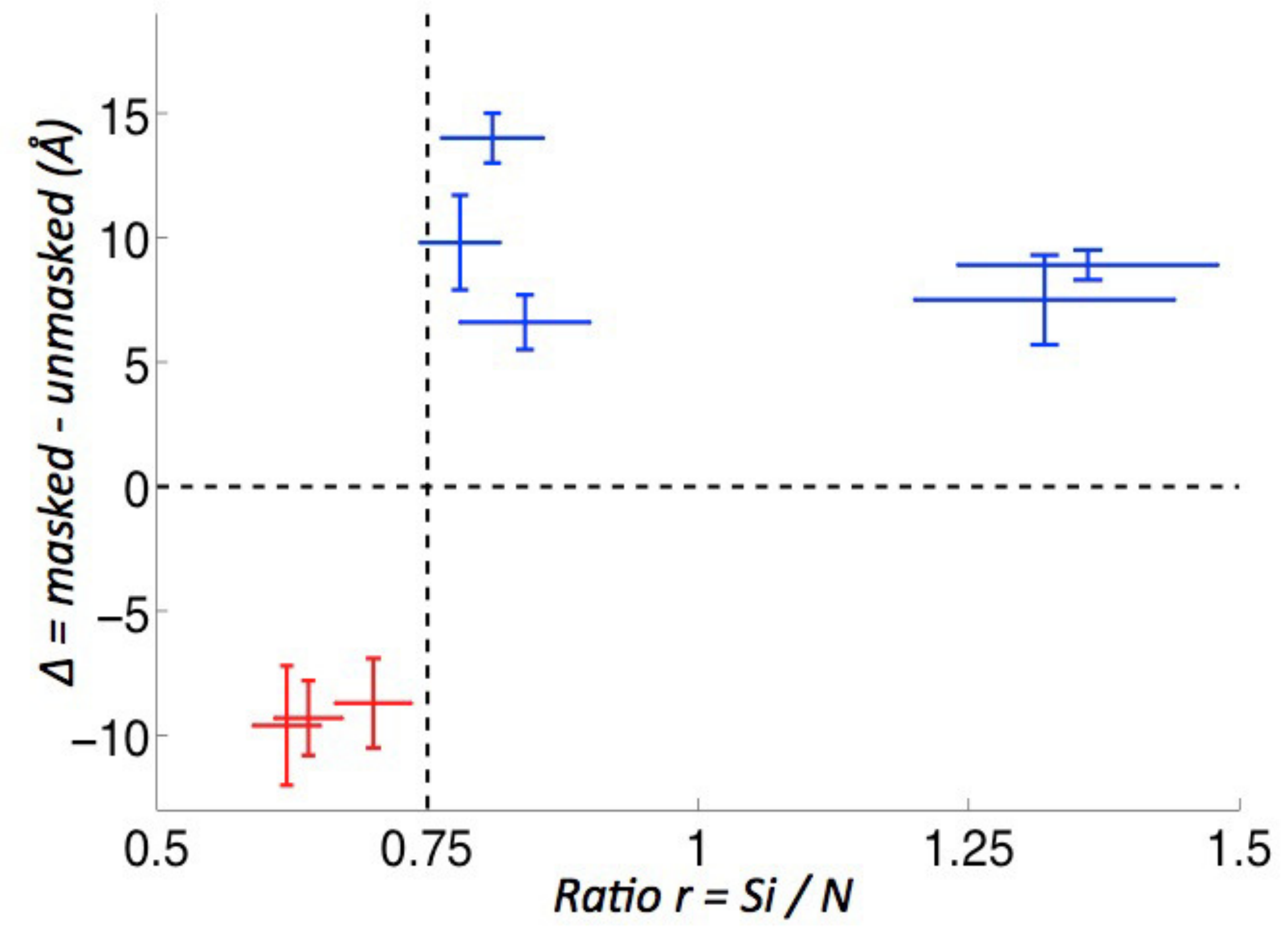}
\end{array}$
\end{center}
\vspace{-18pt}
\caption{Local control of thickness as a function of the composition of the silicon nitride film, of stoichiometric formula Si$_3$N$_4$. In blue, Si-rich silicon nitride mask. In red, the mask was N-rich.}
\label{fig:thickness}
\vspace{-5pt}
\end{figure}

After preparation of the surface of 3.5 x 4.5 mm$^2$ 4H-SiC wafer dies by a high temperature hydrogen etch\cite{deheer2011_3}, silicon nitride is deposited by low-power plasma-enhanced chemical vapor deposition (PECVD) using SiH$_4$ and NH$_3$ as precursor gases. In Fig.\ref{fig:thickness}, a hard mask (glass slide) covers half the SiC die so that the evaporated SiN layer covers only half the sample. We have confirmed by AFM measurements after removing SiN with hydrofluoric acid (HF) that the plasma does not result in detectable damage to the SiC surface. Patterns of SiN were achieved by standard lithography, using PMMA as the resist, and HF as the etchant (harmless to SiC).

\begin{figure}[h]
\vspace{-5pt}
\includegraphics[width=.7\linewidth]{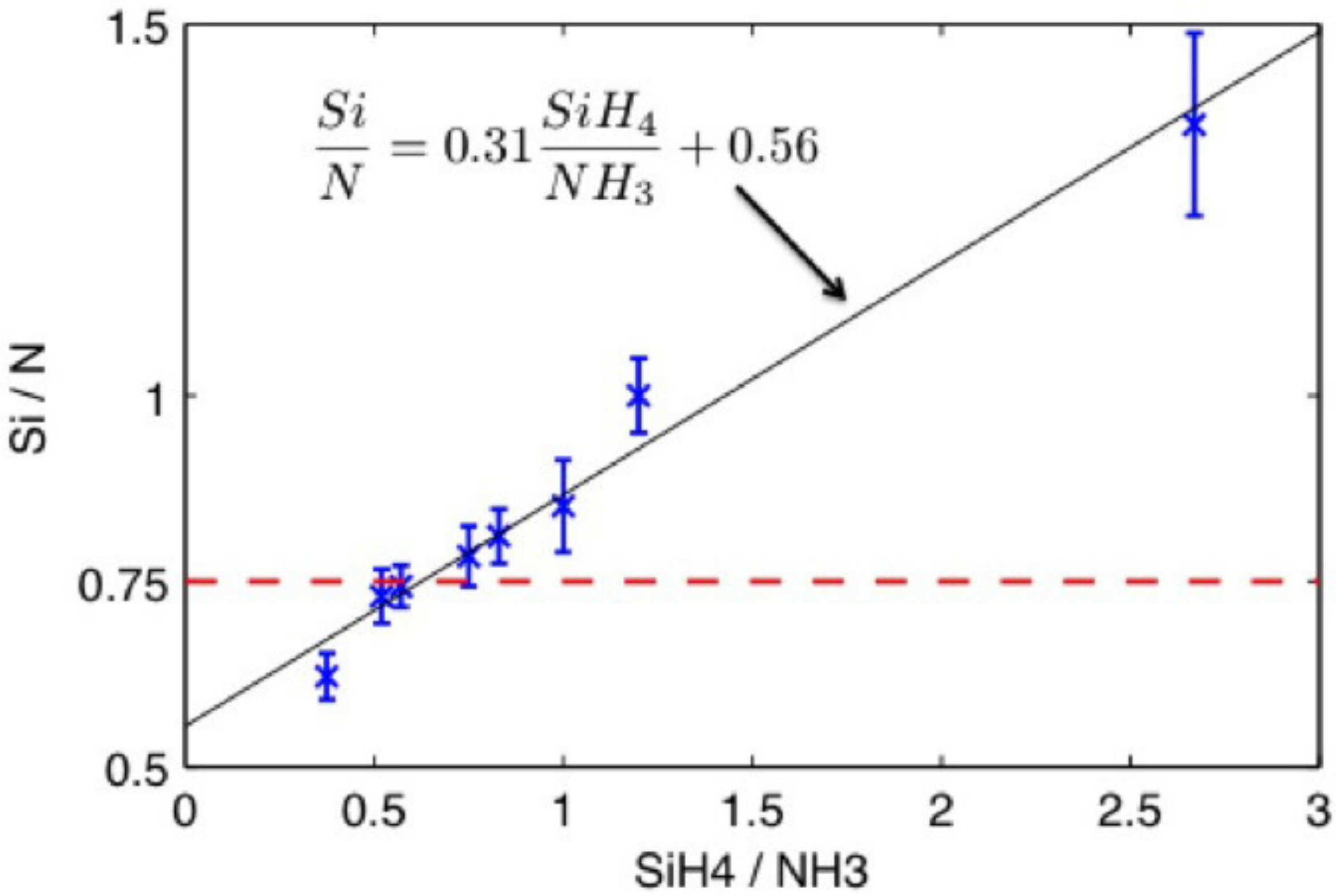}
\vspace{-10pt}
\caption{Calculated ratio of Si over N in the grown SiN film as a function of SiH$_4$/ NH$_3$ precursor ratio used in the PECVD reactor. 0.75 corresponds to stoichiometric Si$_3$N$_4$.}
\label{fig:sin}
\vspace{-10pt}
\end{figure}

We estimated the stoichiometry of the SiN films as a function of the precursor ratio (SiH$_4$ / NH$_3$) by measuring  the  refractive index at 632.8 nm ($n_{632.8}$) of the SiN films with an ellipsometer. Following Refs. \cite{piggins1987,claassen1983}, $n_{632.8}$ is approximately linearly dependent on the ratio Si/N in the deposited film:
	\vspace{-4pt}
	\begin{equation}
	\frac{Si}{N} = \frac{n_{632.8} - 1.35}{0.74}	
	\vspace{-4pt}
	\end{equation}
The plot of  Fig.\ref{fig:sin} gives a calibration of the SiN composition as a function of the precursor ratio (SiH$_4$ / NH$_3$)  in the PECVD process. The straight line is a linear fit of the data. 

The confinement-controlled sublimation (CCS) growth method is used for graphitization\cite{deheer2011_3}. 
This technique consists in heating a  SiC chip in a  graphite enclosure connected to a vacuum chamber (about $10^{-5}$ mbar) by a calibrated hole. This increases the built-in Si partial pressure, which controls the rate of silicon sublimation from the SiC surface, bringing the graphene growth process close to equilibrium. 

The growth process consists of 10 minutes at 800$^\circ$C, followed by graphene growth between 1450 to 1550 $^\circ$C for 8 to 20 minutes. One exception in Fig.\ref{fig:thickness} is sample Si4, which after the 800$^\circ$C annealing, was held at 1150 $^\circ$C for 20 minutes, and then graphitized.  After graphitization the SiN mask has vanished.

Ellipsometry measurements (Horriba Jobin-Yvon AutoSE) on half-masked samples are reported in Fig.\ref{fig:thickness}. We used a  spot size is 250 x 250 \textmu m$^2$,  and analyzed response in the range  440~nm to 850~nm, with a three-term Cauchy model optimized for 4H-SiC and graphene layers \cite{ToBePublished}. Each thickness reported in Fig.\ref{fig:thickness} is the average of 12 measurements, spread on the whole analysed surface.  We observe 2 to 3 additional layers of graphene under the Si-rich initially masked (IM) areas, and consistently 3 fewer layers under the N-rich IM areas, compared to the initially bare (IB) half on each sample. Sample Si4 (Si-rich SiN mask), which had an additional  higher temperature annealing step at 1150$^\circ$C, shows 4 to 5 additional graphene layers.

\begin{figure}[h]
\centering
\vspace{-5pt}
\begin{center}$
\begin{array}{cc}
\includegraphics[width=.9\linewidth]{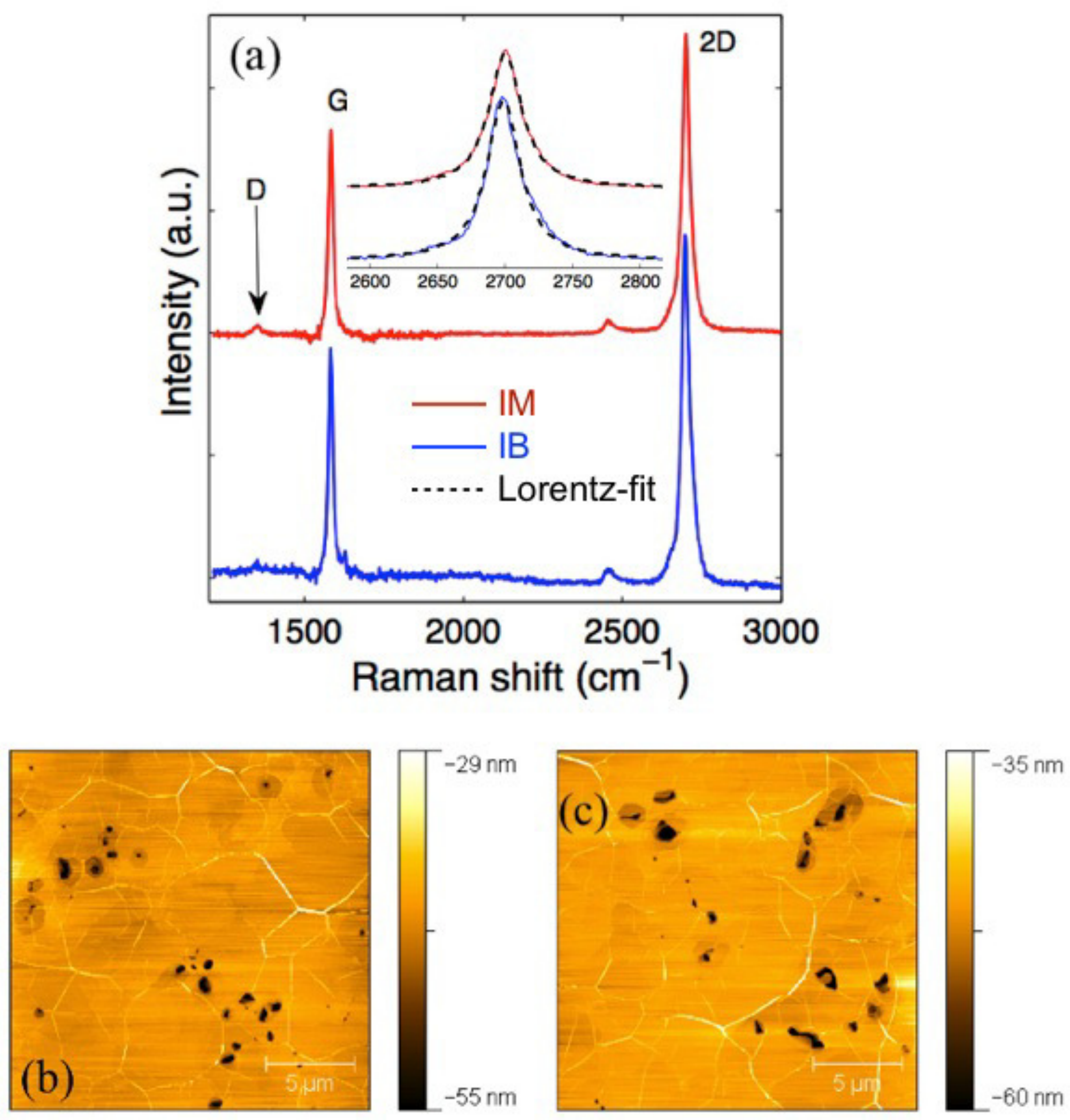}
\end{array}$
\end{center}
\vspace{-18pt}
\caption{Sample N1, N-rich SiN mask. (a): Raman spectra (SiC contribution subtracted) of IM and IB areas.  (b)-(c): AFM images of (b)  IM area and (c) IB area  (scale 20x20 \textmu m$^2$) showing the typical graphene pleat structure observed in MEG samples.}
\label{fig:n12}
\vspace{-5pt}
\end{figure}

\begin{figure}[h!]
\vspace{-5pt}
\begin{center}$
\begin{array}{cc}
\hspace{-5pt}
\includegraphics[width=.95\linewidth]{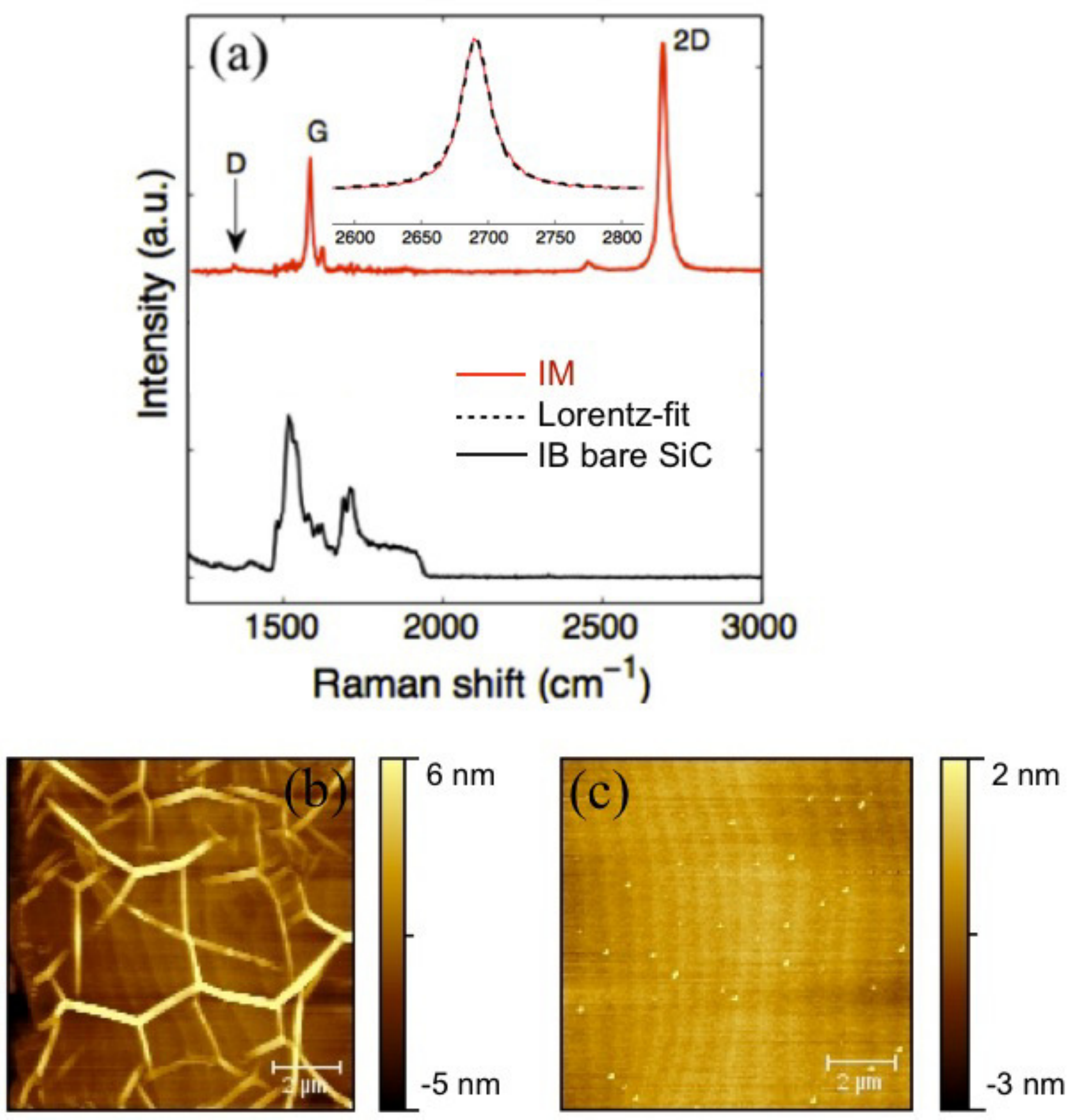}
\end{array}$
\end{center}
\vspace{-15pt}
\caption{ Sample Si1, Si-rich SiN mask. (a): Raman spectra (SiC contribution subtracted), showing the typical MEG spectrum. For the Raman spectrum the intensity is normalized to the SiC plateau at 1900 cm$^{-1}$. Note the quasi absence of D-peak. (b)-(c): AFM image of (b) IM area (scale 10x10 \textmu m$^2$), and (c) IB area (scale 10x10 \textmu m$^2$). }
\label{fig:si15}
\vspace{-5pt}
\end{figure}

Raman spectra (wavelength is 532 nm) and AFM topography images of samples N1 and Si1 are shown in Figs.~\ref{fig:n12} and \ref{fig:si15}, respectively.
Of all the samples studied, the silicon nitride films deposited on N1 and Si1 were the closest to stoichiometric Si$_3$N$_4$, cf Fig.\ref{fig:thickness}. These samples had the lowest Raman D peaks, sharpest 2D peaks, and smoothest  AFM images, and hence were patterned for electronic measurements.

The Raman spectra of the IM and IB areas of N1 and the IM area of Si1, reveal the characteristic graphene peaks, see Figs.~\ref{fig:n12} and \ref{fig:si15}. The graphene 2D and G Raman peaks are clearly identified (the SiC Raman contribution was subtracted). The 2D peak can be fitted by a single Cauchy-Lorentz distribution\cite{faugeras2008} centered at 2699~cm$^{-1}$ for the IM area of N1 (2700~cm$^{-1}$ and 2689~cm$^{-1}$, respectively for the IB part of N1 and the IM half of Si1) with FWHM= 29~cm$^{-1}$ (36~cm$^{-1}$ and 29~cm$^{-1}$, respectively).
 The D peak at 1350 cm$^{-1}$ is  very small, and even undetectable for Si1 and N1. This indicates low defect density in the graphene lattice. For the other samples of Table 1,  the 2D peak is centered  from 2706 to 2726~cm$^{-1}$, with FWHM from 52 to 70 cm$^{-1}$, consistent with 2D MEG \cite{faugeras2008,roehrl2008}. The higher D peaks and the broader blue-shifted 2D peaks (at FWHM 52-70 cm$^{-1}$) for the other samples referenced in Table 1 reveal respectively smaller domain sizes and compressive strain in the graphene \cite{roehrl2008}. Specifically, we do not observed the characteristic shouldered 2D peak of highly ordered pyrolitic graphite (HOPG), as already reported for multilayered epitaxial graphene on the C-face  \cite{faugeras2008,sharma2010,roehrl2008}. The slight asymmetry of the IB 2D peak in Fig.\ref{fig:n12}(a) may be due to a variation of strain in the graphene stack or a small fraction of Bernal stacking fault \cite{faugeras2008}. MEG and FLG on SiC with a 2D-FWHM of 58 cm$^{-1}$ (ref.\cite{zaman2010}), 68 cm$^{-1}$ (ref.\cite{sharma2010}) and even 71 cm$^{-1}$ (ref.\cite{roehrl2008}) have already been reported.
 
The AFM images of Figs.~\ref{fig:n12} and \ref{fig:si15} confirm the presence of graphene, as shown by  the MEG characteristic pleat structure\cite{deheer2011_3}.  In the graphitized areas, the AFM images in  Fig.~\ref{fig:n12}(b)(c) and Fig.~\ref{fig:si15}(b) have a comparable  characteristics in terms of pleat structure, including  pleat height (1.5-2.4 nm),  pleat surface density and semi hexagonal orientation.
Sample Si1 is particularly interesting in that while the  IM area is fully coated with graphene with an average of 3 layers, the IB area  is essentially not graphitized. On the AFM image of Fig.~\ref{fig:si15}(c), a bare SiC  step structure  can be observed, which is confirmed by Raman spectroscopy, cf Fig.\ref{fig:si15}(a). MEG growth is observed in some spots, most probably initiated at screw dislocations in SiC, as already observed \cite{hu2012,gaskill2011}. This indicates  selective growth, with graphene where the mask was, and almost no graphene elsewhere.

\begin{figure}[h!]
\vspace{-5pt}
\includegraphics[width=1.0\linewidth]{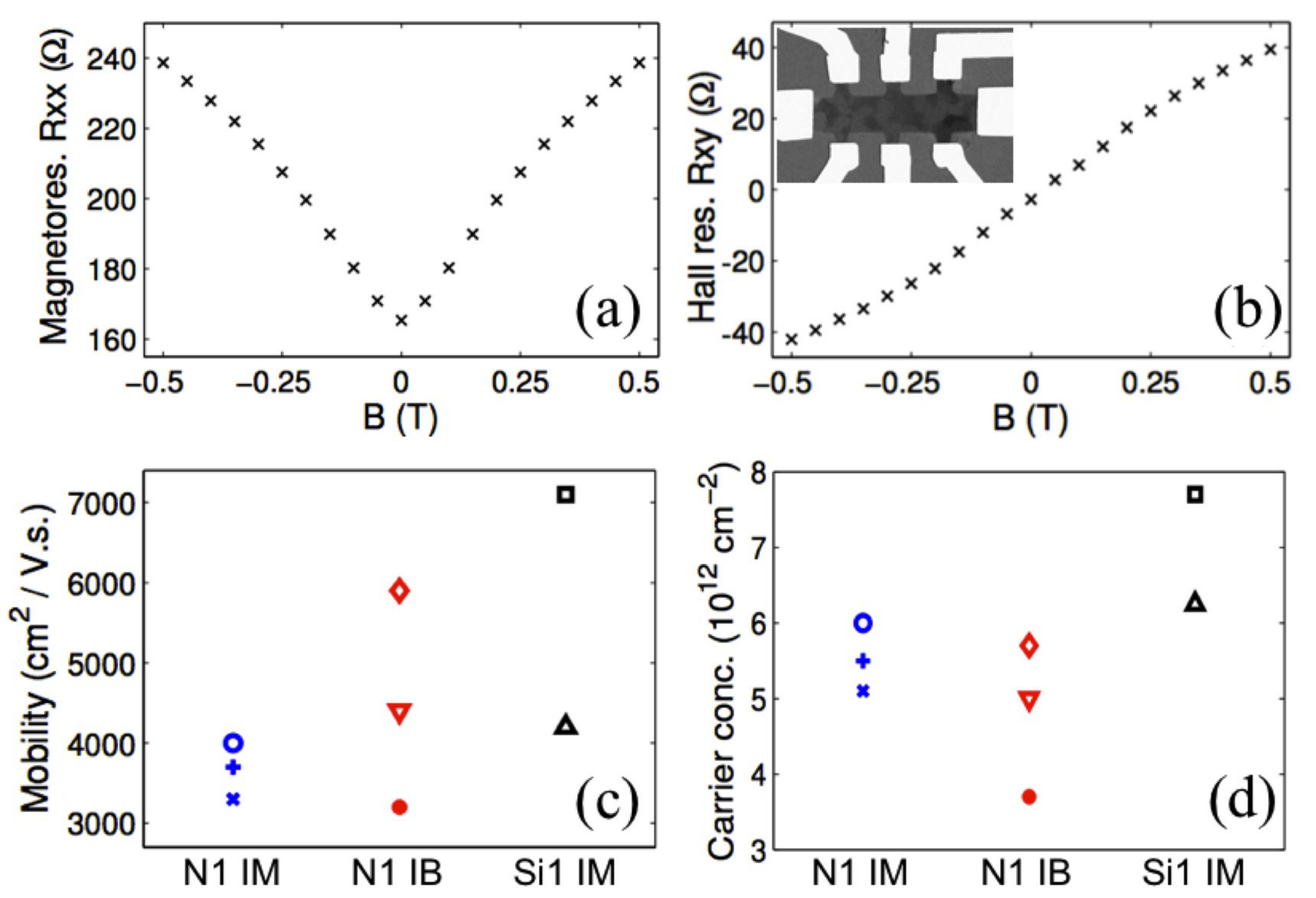}
\vspace{-15pt}
\caption{
	(a) (b): Magneto- and Hall resistances at room temperature of IM area of Si1: n=7.7\ e$^{12}$ cm$^{-2}$ and \textmu~=~7100\ cm$^2$.V$^{-1}$.s$^{-1}$. Hall bar (SEM picture in the inset) is 3~\textmu m wide. (c) (d): Mobility and carrier concentrations at room temperature measured on the IM area of N1, IB area of N1, and IM area of Si1.}
\label{fig:all_transp}
\vspace{-0pt}
\end{figure}

Hall bars (5 \textmu m long, 3.5 \textmu m wide) were patterned in sample Si1 and N1 using electron beam lithography,  oxygen plasma etching and  Ti/Pd/Au contacts (thickness 0.5/20/40 nm). From room temperature Hall and magnetoresistance measurements (Fig.\ref{fig:all_transp}), electronic mobilities are found between 3200 and 7100 cm$^2$.V$^{-1}$.s$^{-1}$, and carrier concentrations are in the $10^{12}~cm^{-2}$ range, showing excellent graphene quality, as seen in Fig.\ref{fig:all_transp}.

\begin{figure}[h!]
\vspace{-5pt}
\begin{center}$
\begin{array}{cc}
\includegraphics[width=.95\linewidth]{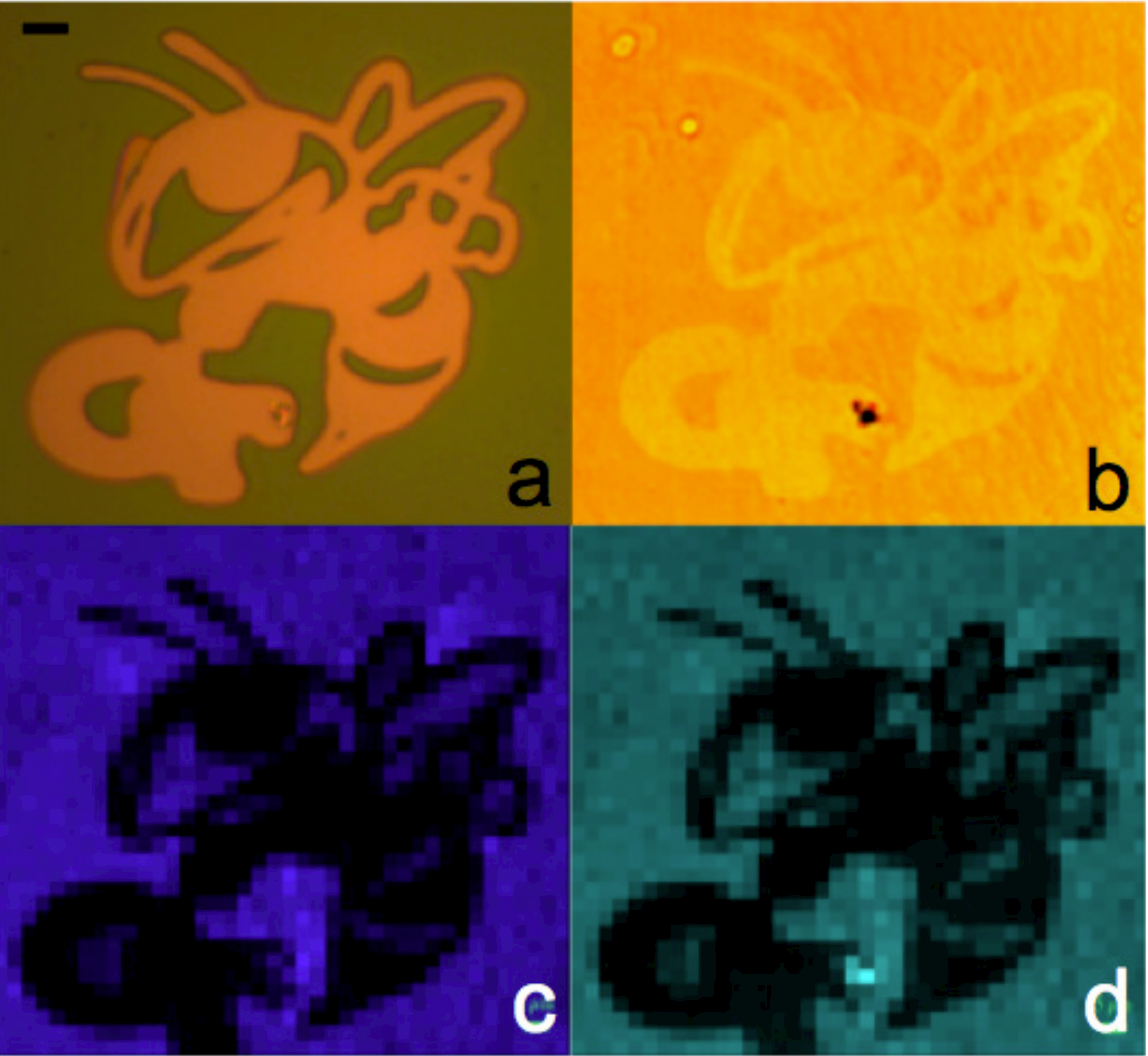} 
\end{array}$
\end{center}
\vspace{-15pt}
\caption{Buzz-of-principle, MEG on SiC using Si-rich SiN mask, demonstrating sub-micron resolution. a: SiN pattern. b: Subsequent MEG growth on SiC, contrast-enhanced optical image. c: Raman 2D map. d: Raman 2D/G map. Scale bar is 10 \textmu m.}
\label{fig:buzz}
\vspace{-10pt}
\end{figure}

As a demonstration of the capability of the method, graphene has been selectively grown in the shape of Buzz, Georgia Tech's mascot. Fig.\ref{fig:buzz} demonstrates that the sub-micrometer resolution of the SiN mask pattern (Fig.\ref{fig:buzz}a) is directly transferred to the selectively grown graphene, as shown by optical contrast (Fig.\ref{fig:buzz}b) and Raman spectroscopy maps of the characteristic 2D and G graphene peaks.  The optical image correlates nicely with the Raman maps, as expected for graphene on SiC \cite{tiberj2011}. The absence of 2D peak in the grey areas of Fig.\ref{fig:buzz}c demonstrates selectivity.


The main result of this study is that  the presence of a  silicon nitride mask evaporated on  silicon carbide prior to graphitization  enhances (Si-rich SiN mask) or reduces (N-rich)  the number of layers grown compared to uncovered areas. Control samples, with no mask, have the same number of layers (within one layer) as the IB areas under the same growth conditions. 

A simple explanation for the reduction of the number of layers under the N-rich masks would be that graphitization is delayed under the mask, starting only after SiN decomposition. It is known that capping SiC with an AlN mask, that doesn't evaporate, prevents graphene formation \cite{zaman2010}. More surprising is the enhanced number of layers under the Si-rich mask.  A possibility is that Si dangling bonds present in the Si-rich-SiN mask \cite{robertson1983,robertson1984,lau1989,warren1993}  react with SiC.  The role of Si in promoting the growth of graphene was demonstrated by implantation of Si in SiC. The implanted SiC surface results in graphene formation at lower temperature than pristine SiC\cite{tongay2012}.  Si dangling bonds at SiN$_3$ sites have already been proposed as the dominant defects in Si-rich and stoichiometric PECVD silicon nitride, acting as amphoteric traps\cite{lau1989}. Moreover, in N-rich films, electron spin resonance (ESR) studies have shown that the density of these defects is greatly reduced or even suppressed\cite{jousse1988}, making N-rich SiN a better dielectric than Si-rich films. However definitive explanations concerning these processes are still premature, as further experimental evidence is necessary. 

We have shown that, by using  a SiN vanishing mask evaporated onto SiC prior to graphitization, the number of graphene layers varies between masked and non-masked areas.  Depending on its chemical composition  (Si- rich or N-rich) the SiN mask acts as an enhancer or inhibitor of graphene growth (+/- 3 graphene layers with the present growth conditions). For few layer  samples, areas with and without graphene can therefore be produced side by side during the heating process. The mask evaporates during graphene growth so that  patterned, mask-free graphene layers are obtained directly in a single heating step. We believe this is a very simple yet potentially quite powerful method to obtain clean patterned graphene structures without the need for post-growth etching.

The authors thank the French Lorraine Region, W. M. Keck Foundation, the National Science Foundation (DMR-0820382), the Air Force Office of Scientific Research for financial support and the Partner University Fund for a travel grant. We also acknowledge funding from the Graphene Flagship European program.
At last my special thanks to J. P. Turmaud, X. Yang, A. Savu, Y. Hu, F. Zaman, S. Bryan, R. Dong, T. Guo, M. Ruan, B. Zhang, M. Sprinkle, J. Hicks, O. Vail, Y. El Gmili, N. Devlin, C. Chapin and D. Brown for training, technical support and fruitful discussions.


\end{document}